\documentclass[10pt,conference]{IEEEtran}
\usepackage{cite}
\usepackage{amsmath,amssymb,amsfonts}
\usepackage{newtxmath}
\usepackage{algorithm}
\usepackage{algpseudocode}
\algnewcommand\algorithmicforeach{\textbf{for each}}
\algdef{S}[FOR]{ForEach}[1]{\algorithmicforeach\ #1\ \algorithmicdo}
\algnewcommand\algorithmicinput{\textbf{Input:}}
\algnewcommand\algorithmicoutput{\textbf{Output:}}
\algnewcommand\Input{\item[\algorithmicinput]}%
\algnewcommand\Output{\item[\algorithmicoutput]}%
\usepackage{tikz}
\usepackage{booktabs}
\usepackage{dcolumn}
\usepackage{ifthen}
\usepackage{listings}
\usepackage{amssymb}
\usepackage{multirow}
\usepackage{pgfplots}
\usepackage{pgfplotstable}
\usepgfplotslibrary{statistics}
\pgfplotsset{compat=1.8}
\usepackage{paralist}
\usetikzlibrary{arrows,automata}
\usetikzlibrary{calc,backgrounds}
\usetikzlibrary{arrows,decorations.markings}
\usepackage{arydshln}

\usepackage{todonotes}

\usepackage{xspace}
\newcommand{\app}{\emph{SCALER}\xspace}

\usepackage[binary-units]{siunitx}
\usepackage{datatool}
\usepackage[english]{fnumprint}

\DTLloaddb{max}{results/summary_values.csv}

\newcommand{\speed}{\DTLfetch{max}{Data}{Prop}{speed}}

\newcommand{\minRecall}{\DTLfetch{max}{Data}{Prop}{recall.min}}
\newcommand{\maxRecall}{\DTLfetch{max}{Data}{Prop}{recall.max}}

\usepackage[pdftex,
]{hyperref}

\begin{document}
\begin{filecontents*}{results/summary_values.csv}
Data,size,speed,specificity.min,specificity.max,recall.min,recall.max
Prop,7,245,-2,-1,25,56
\end{filecontents*}

\title{Scalable Inference of System-level Models \newline from Component Logs}

\author{
\IEEEauthorblockN{
Donghwan Shin\IEEEauthorrefmark{1},
Salma Messaoudi\IEEEauthorrefmark{1},
Domenico Bianculli\IEEEauthorrefmark{1},
Annibale Panichella\IEEEauthorrefmark{1},
Lionel Briand\IEEEauthorrefmark{1},
\\and Raimondas Sasnauskas\IEEEauthorrefmark{2}}

\IEEEauthorblockA{\IEEEauthorrefmark{1}University of Luxembourg, Luxembourg}
\IEEEauthorblockA{\IEEEauthorrefmark{2}SES, Luxembourg}
}

\maketitle

\begin{abstract}
Behavioral software models play a key role in many software
engineering tasks; unfortunately, these models either are not
available during software development or, if available, they quickly
become outdated as the implementations evolve. Model inference
techniques have been proposed as a viable solution to extract finite
state models from execution logs. However, existing techniques do not
scale well when processing very large logs, such as system-level logs
obtained by combining component-level logs. Furthermore, in the case
of component-based systems, existing techniques assume to know the
definitions of communication channels between components. However,
this detailed information is usually not available in the case of systems
integrating 3rd-party components with limited documentation.

In this paper, we address the scalability problem of inferring the
model of a component-based system from the individual component-level
logs, when the only available information about the system are
high-level architecture dependencies among components and a (possibly incomplete) list of
log message templates denoting communication events between components.  Our
model inference technique, called SCALER, follows a divide and conquer
approach. The idea is to first infer a model of each system component
from the corresponding logs; then, the individual component models are
merged together taking into account the dependencies among components,
as reflected in the logs.  We evaluated SCALER in terms of scalability
and accuracy, using a dataset of logs from an industrial system; the
results show that SCALER can process much larger logs than a
state-of-the-art tool, while yielding more accurate models.

\end{abstract}



\section{Introduction}\label{sec:intro}

Behavior models of software system components play a key role in many
software engineering tasks, such as program
comprehension~\cite{Cook:1998:287001}, test case
generation~\cite{6200086}, and model
checking~\cite{clarke2018model}. Unfortunately, such models either are
scarce during software development or, if available, they quickly become
outdated as the implementations evolve, because of the time and cost
involved in generating and maintaining them~\cite{Walkinshaw2010af}.

One possible way to overcome the lack of software models is to use
\emph{model inference} techniques, which extract models---typically in
the form of (some type of) Finite State Machine (FSM)---from execution
logs.  Although the problem of inferring a minimal FSM is
NP-complete~\cite{biermann1972synthesis}, there have been several
proposals of polynomial-time approximation algorithms to infer
FSMs~\cite{biermann1972synthesis,Beschastnikh:2011:LEI:2025113.2025151,LUO201713}
or richer variants, such as gFSM (guarded
FSM)~\cite{walkinshaw2016inferring,mariani2017gk} and gFSM extended
with transition probabilities~\cite{Emam:2018:IEP:3208361.3196883}, to
obtain more faithful models.

Although the aforementioned model inference techniques are fast and
accurate enough for relatively small  programs, all of them
suffer from scalability issues, due to the intrinsic computational
complexity of the problem. This leads to out-of-memory errors or
extremely long, unpractical execution time when processing very large
logs~\cite{wang2016scalable}, such as system-level logs obtained by
combining (e.g., through linearization) component-level logs.  A
recent proposal~\cite{LUO201713} addresses the scalability issue using
a distributed FSM inference approach based on
MapReduce. However, this approach requires to encode the data to be
exchanged between mappers and reducers in the form of key-value
pairs. Such encoding is application-specific; hence, it cannot be used
in contexts---like the one in which this work has been performed---in
which the system is treated as a black-box (i.e., the source code is not available), 
with limited information about the data recorded in the individual components logs.

Another limitation of state-of-the-art techniques is
that they cannot infer, from com\-ponent-level logs, a system-level
model that captures both the individual behaviors of the system's
components and the interactions among them.  Such a scenario can be
handled with existing model inference techniques for distributed
systems, such as CSight~\cite{Beschastnikh:2014:IMC:2568225.2568246},
which typically assume the availability of channels definitions, i.e.,
the exact definition of which events communicate with each other between components. However, this
information is not available in many practical contexts, where the system is
composed of heterogenous, 3rd-party components, with limited documentation 
about the messages exchanged between
components and the events recorded in logs.

In this paper, we address the scalability problem of inferring the
model of a component-based system from the individual component-level
logs (possibly coming from multiple executions), when the only
available information about the system are high-level architecture
dependencies among components and a (possibly incomplete) list of log message templates denoting
communication events between components.  Our goal is to infer a
system-level model that captures not only the components' behaviors
reflected in the logs but also the interactions among them.

Our approach, called \app, follows a \emph{divide and conquer}
strategy: we first infer a model of each component from the
corresponding logs using a state-of-the-art model inference technique,
and then we ``stitch'' (i.e., we do a peculiar type of merge) the
individual component models into a system-level model by taking into
account the dependencies among the components, \emph{as reflected in
  the logs}. The rationale behind this idea is that, though existing
model inference techniques cannot deal with the size of all combined
component logs, they can still be used to infer the models of
individual components, since their logs are sufficiently small. In
other words, \app tames the scalability issues of existing techniques
by applying them on the smaller scope defined by component-level
logs. 

We implemented \app in a prototype tool, which uses
MINT~\cite{walkinshaw2016inferring}, a state-of-the-art technique for
inferring gFSM, to infer the individual component-level models. We
evaluate the scalability (in terms of execution time) and the accuracy
(in terms of recall and specificity) of \app in comparison with MINT
(fed with system-level logs reconstructed from component-level logs),
on \fnumprint{7} proprietary datasets from one of our industrial
partners in the satellite domain. The results show that our approach
\app is about \speed\space times (on average) faster and can process larger
logs than MINT. It generates nearly correct (with specificity always
higher than 0.96) and largely complete models (with an average recall
of 0.79), achieving higher recall than MINT (with a difference ranging between
+\minRecall\thinspace $pp$ and +\maxRecall\thinspace $pp$, with
$pp$=percentage points) while retaining similar specificity.

To summarize, the main contributions of this paper are:
\begin{itemize}
\item the \app approach for taming the scalability problem of
  inferring the model of a component-based system from the individual
  component-level logs, especially when only limited information about
  the system is available;
\item a publicly available implementation of
  \app\footnote{The open-source license is currently being reviewed by our legal team.};
\item the empirical evaluation, in terms of scalability and accuracy,
  of \app and its comparison with a state-of-the-art approach.
\end{itemize}

The rest of the paper is organized as
follows. Section~\ref{sec:background} gives the basic definitions of
logs and models that will be used throughout the
paper. Section~\ref{sec:example} illustrates the motivating example.
Section~\ref{sec:inference} describes the different steps of the core
algorithm of \app. Section~\ref{sec:evaluation} reports on the evaluation of
\app.  Section~\ref{sec:related-work} discusses related work.
Section~\ref{sec:conclusion} concludes the paper and provides
directions for future work.


\section{Background}\label{sec:background}
This section provides the basic definitions for the main concepts that will be 
used throughout the paper.

\paragraph*{Logs} A log is a sequence of log entries; a log entry contains a timestamp
(recording the time at which the logged event occurred) and a log
message (with run-time information related to the logged
event). A log message is a block of free-form text that can be further
decomposed~\cite{messaoudi2018search} into a fixed part called event
template, characterizing the event type, and a variable part, which
contains tokens filled at run time with the values of the event
parameters. For example, given the log entry
\fbox{\texttt{20181119:14:26:00 \textbf{send} OK \textbf{to}
    comp1}}, the timestamp is \texttt{20181119:14:26:00}, the event
template contains the fixed words \texttt{\textbf{send}} and
\texttt{\textbf{to}}, while the tokens \texttt{OK} and \texttt{comp1}
are the values of the event parameters. More formally, let $\mathit{ET}$ be the
set of all events that can occur in a system and $V$ be the set of all
mappings from events parameters to their concrete values, for all
events $\mathit{et}\in \mathit{ET}$; a log $L$ is a sequence of log entries
$\langle e_1, \dots, e_n \rangle$, with $e_i = (\mathit{ts}_i, \mathit{et}_i, v_i)$,
$\mathit{ts}_i\in \mathbb{N}$, $\mathit{et}_i\in \mathit{ET}$, and
$v_i\in V$, for $i=1,\dots,n$. We denote the log of a component
$c_X$ with $L_{c_X}$.
To denote individual log entries, we use the notation $e_{i,j}^{k}$
for the $i$-th log entry of component $k$ in the $j$-th execution; we
drop the subscript $j$ when it is clear from the context.

\paragraph*{Guarded Finite State Machines} We represent the models inferred for a system as guarded Finite State
Machines (gFSMs). A gFSM is a tuple $m=(S, \mathit{ET}, G, \delta, s_0, F)$,
where $S$ is a finite set of states, $\mathit{ET}$ is the set of system events
defined above, $G$ is a finite set of guard functions of the form
$g\colon V\to \{0, 1\}$, $\delta$ is the transition relation
$\delta \subseteq S\times \mathit{ET} \times G \times S$, $s_0\in S$ is the
initial state, $F\subseteq S$ is the set of final states. Informally,
a gFSM is a finite state machine whose transitions are triggered by
the occurrence of an event and are guarded by a function that
evaluates the values of the event parameters. More specifically, a
gFSM $m$ makes a guarded transition from a state $s\in S$ to a
state $s^\prime\in S$ when reading an input log entry
$e=(\mathit{ts},\mathit{et},v)$, written as $s \xrightarrow{e} s^\prime$,
if $(s,\mathit{et},g,s^\prime) \in \delta $ and $g(v)=1$. We say that $m$ \emph{accepts} a log
$l = \langle e_1, \dots, e_n \rangle$ if
there exists a sequence of states
$\langle \gamma_0, \dots, \gamma_n \rangle$ such that (1)
$\gamma_i \in S$ for $i=0,\dots,n$, (2) $\gamma_0 = s_0$, (3)
$\gamma_{i-1} \xrightarrow{e_i} \gamma_{i}$ for
$i=1,\dots,n$, and (4) $\gamma_n \in F$.

\section{Motivations}\label{sec:example}
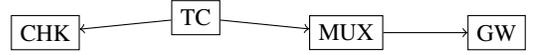
\begin{figure}[tb]
 \centering
 \begin{tikzpicture}
    \node[draw] (tc) at (0,0) {\small TC};
    \node[draw] (mux) at (2,-0.2) {\small MUX};
    \node[draw] (chk) at (-2,-0.2) {\small CHK};
    \node[draw] (gw) at (4,-0.2) {\small GW};

    \draw[->] (tc) to (mux);
    \draw[->] (tc) to (chk);
    \draw[->] (mux) to (gw);
  \end{tikzpicture}
 \caption{The components of the example system and their dependencies}
 \label{fig:components}
\end{figure}

In this section, we discuss the motivations for this work using an
example based on a real system from one of our industrial partners in the
satellite domain. We consider a simplified version of a satellite
ground control system, composed of the four components shown in
Figure~\ref{fig:components}: \textit{TC}, the module handling
tele-commands for the satellite, which is also the entry point of the
system; \textit{MUX}, a multiplexer combining different tele-commands
into a single communication stream; \textit{CHK}, the module
validating the tele-commands parameters before they are sent to the
satellite; \textit{GW}, the gateway managing the connections between
the satellite and the ground control system.
Figure~\ref{fig:components} also shows the architectural dependencies
among components; for example, the arrow from component \textit{TC} to
component \textit{MUX} indicates that \textit{TC} \emph{uses} (or
invokes) an operation provided by \textit{MUX}.  Every execution of
the system generates a set of logs, with one log for each component;
Figure~\ref{fig:logs} depicts the logs
of the four system components generated in two executions; for space
reasons, the format of timestamps has been compressed.

\begin{figure}
\renewcommand{\arraystretch}{1.3}
 \resizebox{9cm}{!}{%
      \begin{tabular}{l|l|l}
      \hline
      \textbf{CMP} & \multicolumn{1}{c|}{\textbf{Execution 1}} & \multicolumn{1}{c}{\textbf{Execution 2}} \\
      \hline
      \multirow{2}{*}{TC}	& $e_{1,1}^{\mathit{TC}}$= \texttt{14:26:01 sending X via f0} & $e_{1,2}^{\mathit{TC}}$= \texttt{14:30:11 sending Y via f1}\\
      					& $e_{2,1}^{\mathit{TC}}$= \texttt{14:26:02 TC accepted} & $e_{2,2}^{\mathit{TC}}$= \texttt{14:30:12 wait message} \\ 
     \hline
      \multirow{5}{*}{MUX}	& $e_{1,1}^{\mathit{MUX}}$= \texttt{14:26:01 initialize} & $e_{1,2}^{\mathit{MUX}}$= \texttt{14:30:11 initialize}  \\
					& $e_{2,1}^{\mathit{MUX}}$= \texttt{14:26:01 commandName = X} & $e_{2,2}^{\mathit{MUX}}$= \texttt{14:30:12 commandName = Y}\\
					& $e_{3,1}^{\mathit{MUX}}$= \texttt{14:26:01 commandName = X} & $e_{3,2}^{\mathit{MUX}}$= \texttt{14:30:12 data flow ID = f1}  \\
					& $e_{4,1}^{\mathit{MUX}}$= \texttt{14:26:01 data flow ID = f0} & $e_{4,2}^{\mathit{MUX}}$= \texttt{14:30:12 send = no} \\
					& $e_{5,1}^{\mathit{MUX}}$= \texttt{14:26:02 send= ok} &  \\	
	\hline
      \multirow{1}{*}{GW}	& $e_{1,1}^{\mathit{GW}}$= \texttt{14:26:01 encrypt TC\_01} & $e_{1,2}^{\mathit{GW}}$= \texttt{14:30:12 reject command}  \\

      \hline
      \multirow{2}{*}{CHK}	& $e_{1,1}^{\mathit{CHK}}$= \texttt{14:26:01 mode 1} & $e_{1,2}^{\mathit{CHK}}$= \texttt{14:30:11 mode 0}  \\ 
      					& $e_{2,1}^{\mathit{CHK}}$= \texttt{14:26:02 automatic config} & \\
\hline				
    \end{tabular}
  }
  \renewcommand{\arraystretch}{1.3}
\centering
  \resizebox{8cm}{!}{
   \begin{tabular}{|l|l|l|}
      \multicolumn{3}{c}{\textbf{Log Message Templates}} \\
      \hline
      $*\mathit{tmp}_1$= sending $v_1$ via $v_2$ & $*\mathit{tmp}_2$= TC accepted & $*\mathit{tmp}_3$= wait message  \\
      $*\mathit{tmp}_4$= initialize & $\mathit{tmp}_5$= cmdName = $v_1$ & $\mathit{tmp}_6$= data flow ID = $v_1$ \\
      $*\mathit{tmp}_7$= send = $v_1$  & $*\mathit{tmp}_8$= encrypt $v_1$ & $*\mathit{tmp}_9$= reject command  \\
      $*\mathit{tmp}_{10}$= mode $v_1$ & $*\mathit{tmp}_{11}$= automatic config &\\
   	\hline
    \end{tabular}
    }
   \caption{(top) Component logs generated by two executions of the example
    system; (bottom) Log message templates  extracted from
    components logs (communication message templates are marked with an asterisk).}
   \label{fig:logs}
\end{figure}

To infer a model from these individual component logs, one could use
existing model inference techniques for distributed systems, such as
CSight~\cite{Beschastnikh:2014:IMC:2568225.2568246}. These techniques
typically assume the availability of channels definitions, i.e., the
exact definition of which events communicate to each other between
components. However, this information is not available in many
practical contexts, including ours, where the system is composed of
heterogenous, 3rd-party components, with limited documentation. More
specifically, the only available information about the system are
high-level architecture dependencies among components (like those in
Figure~\ref{fig:components}) and a (possibly incomplete) list of
communication events, without knowing exactly how events communicate with
each other. Due to this limited information, we cannot use existing
techniques for model inference for distributed systems.

Another approach towards model inference would be to reconstruct a
system-level log from the individual component logs and use
non-distributed model inference techniques such as
MINT~\cite{walkinshaw2016inferring} or GK-tail+~\cite{mariani2017gk}.
However, such approaches typically suffer from scalability issues due to
the underlying algorithms they use. For example, the main algorithm
used in MINT has worst-case time complexity that is cubic in the size of the
inferred model~\cite{10.1007/BFb0054059}; the algorithm used for
removing non-determinism from models can exhibit, based on our
preliminary evaluation, deep recursion that causes stack
overflows and makes MINT crash.
Furthermore, GK-tail+ is not publicly
available and the largest log on which it was evaluated contained
11386 log entries. Since the system of our industrial partner
can generate, when considering all the components, logs with more than
30000 entries, there is need for a scalable model inference
technique that can process  component logs.

\section{Scalable Model Inference}\label{sec:inference}

Our technique for system model inference from component logs follows a
\emph{divide and conquer} approach. The idea is to first infer a model
of each system component from the corresponding logs; then, the
individual component models are merged together taking into account
the dependencies among components, \emph{as reflected in the logs}. We
call this process \app. The rationale behind our
technique is that though existing (log-based) model inference
techniques cannot deal with the size of all combined component logs,
they can still be used to accurately infer the models of individual components,
since their logs are sufficiently small for the existing model inference 
techniques to work. The challenge is then how to
``stitch'' together the models of the individual components to build a
system model that reflects not only the components behavior but also
their dependencies, while preserving the accuracy of the component models.
For example, simply appending one component model after the other
perfectly preserves the accuracy of the inferred component models,
but it significantly loses the dependencies between components.
On the other hand, performing a parallel composition of automata 
on the component models (based on the dependencies between components) 
loses the accuracy of the component models because of 
the over-generalization caused by the parallel composition.
To solve this problem, we develop a set of novel algorithms that 
take into account the dependencies between components
while preserving the component models as much as possible.

\begin{figure}
	\centering
	\begin{tikzpicture}
\tikzset{myblock/.style = {rectangle, draw, text width=17mm, minimum height=2.5cm}}
    
    \node[text width=2cm, text centered] at (-1, 1.5) {\footnotesize Preprocessing};
    \node[text width=2cm, text centered] at (-3, 0.8) {\scriptsize Individual\\[-5pt] components logs};
    \node[text width=2cm, text centered] at (-3, -0.3) {\scriptsize
      Architectural\\[-5pt] dependencies};

    \node[text width=2cm, text centered] at (-3, -0.95) {\scriptsize
      Communication\\[-5pt] events templates};
    
    \node (foo)[myblock] at (-1, 0) {};
    \draw[->] (-2.1, 0.8) -- (-1.8, 0.8);
    \draw[->] (-2.1, -0.9) -- (-1.8, -0.7);
    \draw[->] (-2.1, -0.3) -- (-1.8, -0.5);
    \node (mint)[myblock] at (-1,0.6) [dotted, text centered, text width=13.5mm, minimum height=0.8cm] {\scriptsize MINT};
    \node (dep)[myblock] at (-1,-0.6) [dotted, text centered, text width=13.5mm, minimum height=1cm] { \scriptsize  Log entries\\[-1pt]dependencies\\[-5pt]extraction};

    \node[text width=2cm, text centered] at (2.8, 1.5) {\footnotesize Stitching};
    
    \node (bar)[myblock,right of=foo,xshift=2.8cm]{};
        \node[text width=1cm, text centered] at (2.8, 1) {\scriptsize Stitch()};
    \node (graft)[myblock] at (2.8, -0.1) [dotted, text width=13mm, minimum height=1.8cm] {};
    	\node[text width=1cm, text centered] at (2.8, 0.6) {\scriptsize Graft()};
    \node (slice)[myblock] at (2.8, 0.1) [dotted, text centered, text width=9mm, minimum height=0.5cm] {\scriptsize Slice()};
    \node (insert)[myblock] at (2.8, -0.6) [dotted, text centered, text width=9mm, minimum height=0.5cm] {\scriptsize Insert()};
    \draw[->] (graft.east) -- +(0.1,0) |- +(0.1,-1.08) -| node {} (graft.south);
    
    \draw[->] (-0.2, 0.6) -- (0.1, 0.6);
     \draw[->] (1.5, 0.6) -- (1.8, 0.6);
    \node[text width=2cm, text centered] at (0.9, 0.6) {\scriptsize Components\\[-5pt] gFSMs};
    
    \draw[->] (-0.2, -0.6) -- (0.1, -0.6);
     \draw[->] (1.5, -0.6) -- (1.8, -0.6);
    \node[text width=2.5cm, text centered] at (0.9, -0.6) {\scriptsize Log entries\\[-5pt] dependencies};
     
     \draw[->] (3.8, 0) -- (4.1, 0);
     \node[text width=1cm, text centered] at (4.4, 0) {\scriptsize System\\[-5pt] gFSM};

\end{tikzpicture}
	\caption{Workflow of the \app technique}
	\label{fig:overview}
\end{figure}
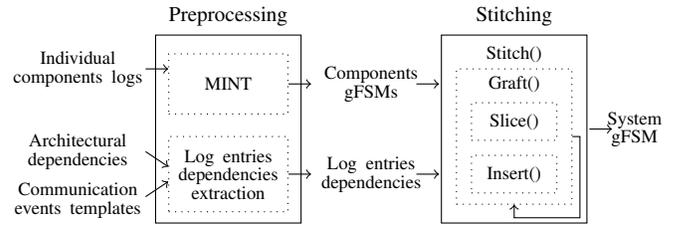

Figure~\ref{fig:overview} outlines the workflow of \app. The technique
takes as input the logs of the different components, possibly coming
from multiple executions, a description of the architectural
dependencies among components, and  a list of log message templates denoting
communication events between components; it returns a system level gFSM. 
\app is composed of two stages, \emph{pre-processing} and
\emph{stitching}, described in the following subsections.

\subsection{Pre-processing Stage}
\label{sec:pre-processing}
This stage prepares two intermediate outputs,
\emph{component-level models} and \emph{log entries dependencies},
which will be used by the main stitching stage.

\subsubsection*{Inferring Component  Models}\label{sec:mint}
For each component, we infer a component-level model
based on the corresponding logs using 
MINT~\cite{walkinshaw2016inferring}, an open-source state-of-the-art tool.

MINT takes as input
\begin{inparaenum}[(1)]
\item  the logs produced by the individual component for which one
wants to infer the model and
\item  the templates of the events recorded in
the component logs.
\end{inparaenum}
The event templates are required to parse the log
entries, to retrieve the actual events and their parameters.
Nevertheless, often such templates are not available or
documented. This situation is typical when dealing with 3rd-party,
black-box components---as it is the case for the ground control system
used by our industrial partner---and it is known in the literature as
the log message format identification problem. We use
MoLFI~\cite{messaoudi2018search}, a state-of-the-art solution for this
problem, to derive the event templates that are then used by MINT; as
an example, the box at the bottom of Figure~\ref{fig:logs} 
shows the templates produced by
MoLFI from the logs of our running example.

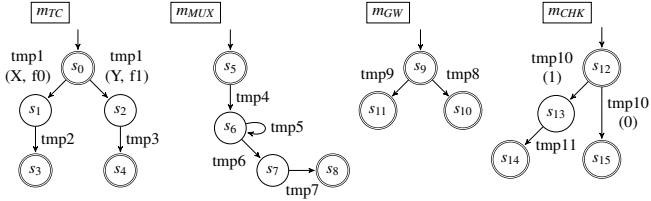
\begin{figure}
	\resizebox{\linewidth}{!}{\tikzset{myblock/.style = {rectangle, draw, text width=25mm, minimum height=3cm}}

\hspace*{0em}{\begin{tikzpicture}[baseline=0pt,->,>=stealth',shorten >=1pt,auto,node distance=1.3cm, initial text = {}, 
	state/.style={circle, draw, minimum size=0.4cm}]
\tikzstyle{every state}=[text=black]
  \node[draw] at (-0.6,1.2) {$m_{\mathit{TC}}$};
  \node[state, initial, accepting, initial where=above] (A) {$s_0$};
  \node[state]         (B) [below left of=A] {$s_1$};
  \node[state]         (C) [below right of=A] {$s_2$};
  \node[state]         (D) [accepting, below of=B] {$s_3$};
  \node[state]         (E) [accepting, below of=C]       {$s_4$};

  \path (A) edge   node[above left, align=center] {tmp1\\(X, f0)} (B)
                 edge              node [align=center] {tmp1\\(Y, f1)} (C)
       	 (B) edge   		node {tmp2} (D)
        	(C) edge              node {tmp3} (E);
\end{tikzpicture}}
\hspace*{0em}{\begin{tikzpicture}[baseline=0pt,->,>=stealth',shorten >=1pt,auto,node distance=1.3cm, initial text = {}, 
state/.style={circle, draw, minimum size=0.4cm}]
  \tikzstyle{every state}=[text=black]
  \node[draw] at (-0.75,1.2) {$m_{\mathit{MUX}}$};
  \node[state, initial, accepting, initial where=above] (x) {$s_5$};       
  \node[state]         (y) [below of=x] {$s_6$};
  \node[state]         (z) [below right of=y] {$s_7$};
  \node[state]         (f) [accepting, right of=z] {$s_8$};

  \path (x) edge       node {tmp4} (y)
       	 (y) edge         node[below left]  {tmp6} (z)
	      edge [loop right] node {tmp5} ()
        	(z) edge              node [below =0.18cm] {tmp7} (f);
\end{tikzpicture}}
\hspace*{0em}{\begin{tikzpicture}[baseline=0pt,->,>=stealth',shorten >=0.7pt,auto,node distance=1.3cm, initial text = {}, 
state/.style={circle, draw, minimum size=0.4cm}]
  \tikzstyle{every state}=[text=black]
\node[draw] at (-0.7,1.2) {$m_{\mathit{GW}}$};
  \node[state, initial, accepting, initial where=above] (x) {$s_9$};
  \node[state]         (y) [accepting, below right of=x] {$s_{10}$};
  \node[state]         (z) [accepting, below left of=x] {$s_{11}$};

  \path (x) edge       node {tmp8} (y)
       	 edge         node [above left] {tmp9} (z);
\end{tikzpicture}}
\hspace*{0em}{\begin{tikzpicture}[baseline=0pt, ->,>=stealth',shorten >=1pt,auto,node distance=1.4cm, initial text = {}, 
state/.style={circle, draw, minimum size=0.4cm}]
  \tikzstyle{every state}=[text=black]
  \node[draw] at (-0.75,1.2) {$m_{\mathit{CHK}}$};
  \node[state, initial, accepting, initial where=above] (x) {$s_{12}$};
  \node[state]         (y) [below left of=x] {$s_{13}$};
  \node[state]         (z) [accepting, below left of=y] {$s_{14}$};
  \node[state]         (f) [accepting, below right of=y] {$s_{15}$};

  \path (x) edge     node  [above left, align=center] {tmp10\\(1)} (y)
       	        edge         node [align=center] {tmp10\\(0)} (f)
	   (y) edge       node [below right =-0.1cm] {tmp11} (z);
\end{tikzpicture}}}
	\caption{Component-level gFSMs inferred by MINT from the logs shown in Table~\ref{fig:logs}}
	\label{fig:component-models}
\end{figure}

The models inferred by MINT are gFSMs;
Figure~\ref{fig:component-models} shows the component-level gFSMs
inferred by MINT for the four components of our running example. We
use a compact notation for the guards on the event
parameters labeling the guarded transitions; for example, in the gFSM of
\textit{TC} (i.e., $m_{\mathit{TC}}$), the guard $(\texttt{X}, \texttt{f0})$ stands for 
$(v_1 = \text{``\texttt{X}''}, v_2 = \text{``\texttt{f0}''})$.

\subsubsection*{Identifying Log Entries Dependencies}\label{sec:causality}

A system-level model of a component-based system has to capture not
only the behavior of the individual components but also the intrinsic
behavioral dependencies among them. For example, considering the fact that 
\textit{TC} invokes \textit{MUX} as shown in Figure~\ref{fig:components}, one 
could speculate that the event recorded in entry $e_{1,1}^{TC}$ could lead to 
the event recorded in entry $e_{1,1}^{MUX}$ in Figure~\ref{fig:logs}; if this is 
the case, the model should reflect this dependency, 
which we call \emph{log entries dependency}.

Log entries dependencies can be extracted from the source code by means
of program analysis or from existing models such as UML Sequence
Diagrams~\cite{Whaley:2002:AEO:566172.566212,1707665}.  
However, the extraction is infeasible when the source code is
not available and the documentation is limited. This is
the case for the example system provided by our industrial partner:
the source code of 3rd-party components is not available, the
architectural documentation only includes coarse-grained dependencies
(like those shown in Figure~\ref{fig:components}), and the only
additional information is the knowledge of domain experts, who can
provide a (possibly incomplete) list of log message templates corresponding to
events related to the ``communication'' between components (like those
marked with an asterisk in Figure~\ref{fig:logs}).
To solve this issue, we present a simple heuristic that identifies
log entries dependencies from the coarse-grained component dependencies,
the list of communication event templates, and the individual component logs.

The idea at the basis of our heuristic is that, if there is an
architectural dependency from a component $c_X$ to another component
$c_Y$ (representing the use of $c_Y$ by $c_X$), then all log entries
of $c_Y$ are ultimately the consequences of the log entries of $c_X$.
The identification of log entries dependencies boils down to
finding out, for each pair of components $c_X$ and $c_Y$ with $c_X$ using
$c_Y$, which log entries of $c_Y$ are the consequences of which log
entry of $c_X$. We observe that, among the log entries of $c_Y$, the communication 
(events recorded in) log entries are invoked directly from $c_X$.
Furthermore, given the timestamp\footnote{We assume that the clocks of the different components 
are synchronized, for example using the Network Time Protocol
(NTP)~\cite{103043}.} of a communication log entry 
$\mathit{ce}^Y$ of $c_Y$, we observe that $\mathit{ce}^Y$ communicated
with  the most recent communication 
log entry $\mathit{ce}^X$ of $c_X$.
Based on these observations, we say that $\mathit{ce}^X$ \emph{(communicatively) 
leads-to} $\mathit{ce}^Y$, denoted with $\mathit{ce}^X \rightsquigarrow_c 
\mathit{ce}^Y$, only if (1) the timestamp of $\mathit{ce}^X$ is less than or 
equal to the one of $\mathit{ce}^Y$ and (2) the timestamp difference between 
$\mathit{ce}^X$ and $\mathit{ce}^Y$ is the minimum among all pairs of the 
communication log entries of $c_X$ and $c_Y$.
In our running example, given the list of templates corresponding to (log 
entries of) communication events: $\mathit{tmp}_1$, $\mathit{tmp}_2$, 
$\mathit{tmp}_4$, and $\mathit{tmp}_7$, if we consider the architectural 
dependency from \textit{TC} to \textit{MUX} and focus on the first execution, we 
say that $e^\mathit{TC}_1 \rightsquigarrow_c e^\mathit{MUX}_1$ and 
$e^\mathit{TC}_2 \rightsquigarrow_c e^\mathit{MUX}_5$.

By definition, the $\rightsquigarrow_c$ relationship does not hold
between the remaining non-communication (events recorded in) log
entries of $c_Y$ and the communication log entries of $c_X$.  However,
since all log entries of $c_Y$ are ultimately the consequence of the
log entries of $c_X$, we can speculate that a sequence of
non-communication log entries
$\langle \mathit{ne}_1^Y, \mathit{ne}_2^Y, \dots, \mathit{ne}_k^Y
\rangle$ of $c_Y$ after a communication log entry $\mathit{ce}^Y$ of
$c_Y$ is also related to the most recent communication log entry
$\mathit{ce}^X$ of $c_X$.  More precisely, if we have a log
$\langle \dots, \mathit{ce}^Y, \mathit{ne}_1^Y, \mathit{ne}_2^Y,
\dots, \mathit{ne}_k^Y, \mathit{ce}'^Y, \dots \rangle$ of $c_Y$ where
$\mathit{ce}^X \rightsquigarrow_c \mathit{ce}^Y$, we say that
$\mathit{ce}^X$ \emph{leads-to}
$\langle \mathit{ce}^Y, \mathit{ne}_1^Y, \mathit{ne}_2^Y, \dots,
\mathit{ne}_k^Y \rangle$, denoted with
$\mathit{ce}^X \rightsquigarrow \langle \mathit{ce}^Y,
\mathit{ne}_1^Y, \mathit{ne}_2^Y, \dots, \mathit{ne}_k^Y \rangle$.
When considering \textit{TC} and \textit{MUX} in the first execution
of our running example, we have
$e^\mathit{TC}_1 \rightsquigarrow \langle e^\mathit{MUX}_1,
e^\mathit{MUX}_2, e^\mathit{MUX}_3, e^\mathit{MUX}_4 \rangle$ because
$e^\mathit{TC}_1 \rightsquigarrow_c e^\mathit{MUX}_1$ (as identified
above); also, we have
$e^\mathit{TC}_2 \rightsquigarrow \langle e^\mathit{MUX}_5 \rangle$
because $e^\mathit{TC}_2 \rightsquigarrow_c e^\mathit{MUX}_5$ and
there are no further non-communication log entries after
$e^\mathit{MUX}_5$.
Table~\ref{tab:causality} shows all the log entries dependencies extracted 
for the log entries in Figure~\ref{fig:logs}.

We remark that our heuristic may introduce some imprecisions,
for example, with logs in which the timestamp granularity is relatively 
coarse-grained (e.g., seconds instead of milli- or nano-seconds) and the 
communication between components is fast enough such that often two 
communication events that logically occur one before the other are logged using 
the same timestamp.
Incorrectly identified log entries dependencies can decrease the accuracy 
of the resulting system-level model and increase its
complexity; we leave the study of more accurate techniques for the identification 
of log entries dependencies as part of future work.

\begin{table}
\centering
\caption{Extracted log entries dependencies for the running example}
\label{tab:causality}
 \resizebox{0.7\linewidth}{!}{%
 \renewcommand{\arraystretch}{1.3}
\begin{tabular}{c|l}
\hline
\textbf{Execution} & \multicolumn{1}{c}{\textbf{Log entry dependencies}} \\ \hline
 \multirow{3}{*}{Exec1} & $e^{\mathit{TC}}_1 \rightsquigarrow  \langle e_1^{\mathit{MUX}}, e_2^{\mathit{MUX}}, e_3^{\mathit{MUX}}, e_4^{\mathit{MUX}} \rangle$ \\
				& $e^{\mathit{TC}}_1 \rightsquigarrow \langle e_1^{\mathit{CHK}} \rangle$ , $e^{\mathit{TC}}_2 \rightsquigarrow \langle e_5^{\mathit{MUX}} \rangle$ \\
				& $e^{\mathit{TC}}_2 \rightsquigarrow \langle e_2^{\mathit{CHK}} \rangle$, $e^{\mathit{MUX}}_4 \rightsquigarrow \langle e_1^{\mathit{GW}} \rangle$ \\ 
\hline
 \multirow{3}{*}{Exec2} & $e^{\mathit{TC}}_1 \rightsquigarrow \langle e_1^{\mathit{MUX}}, e_2^{\mathit{MUX}}, e_3^{\mathit{MUX}} \rangle$ \\
				& $e^{\mathit{TC}}_1 \rightsquigarrow \langle e_1^{\mathit{CHK}} \rangle$ , $e^{\mathit{TC}}_2 \rightsquigarrow \langle e_4^{\mathit{MUX}} \rangle$ \\
				& $e^{\mathit{MUX}}_4 \rightsquigarrow \langle e_1^{\mathit{GW}} \rangle$ \\ 
\hline
\end{tabular}}
\end{table}
\subsection{Stitching Stage}\label{sec:algorithms}

The intermediate outputs of the pre-processing stage are then used in
this \emph{stitching} stage, which is at the core of our technique.
In this stage, we build a system-level gFSM that captures not only the 
components' behavior inferred from the logs but also their
dependencies as reflected in the log entries dependencies identified
in the pre-processing stage.

Since the dependencies between components observed through the logs 
are different from execution to execution,
we first build system-level gFSM \emph{for each execution} and then merge
these gFSMs together using the standard DFA (Deterministic Finite
Automaton) union operation\footnote{MINT produces a deterministic gFSM
  $m=(S, \mathit{ET}, G, \delta, s_0, F)$, with
  $\delta: S\times \mathit{ET}\times G\to S$; it can be easily
  converted into a DFA $m'=(S, \Sigma, \delta', s_0, F)$ with
  $\delta': S\times \Sigma \to S$ where
  $\Sigma = \mathit{ET}\times G$.}. 
We call this process ``stitching'' whereas we call ``grafting'' the inner process that builds 
a system-level gFSM for each execution.
The pseudocode of the top-level process \textsc{Stitch} 
is shown in Algorithm~\ref{fig:stitch}.
\begin{algorithm}[t]
\footnotesize
\begin{algorithmic}[1]
\Input{Set of Components $C = \{c_\mathit{main}, c_1, \dots, c_n\}$ \newline
 \phantom{Seti}Set of gFSMs $M = \{m_{c_{\mathit{main}}}, m_{c_1}, \dots, m_{c_n}\}$ \newline
 \phantom{Seti}Set of Logs $L_{\mathit{main}} = \{l_1, \dots, l_k\}$}
\Output{System model $m_{\mathit{sys}}$}
    \State Set of gFSMs $W \gets \emptyset$ \label{alg:stitch:beginW}
    \ForEach{$l_i \in L_{\mathit{main}}$}
        \State gFSM $m_\mathit{main} \gets$ \textsc{Graft}($c_{\mathit{main}}, l_i, M$)
        \State $W \gets  \{m_\mathit{main}\} \cup W$ \label{alg:stitch:endW}
    \EndFor
    \State gFSM $m_{\mathit{sys}} \gets \mathit{DFAUnion}(W)$ \label{alg:stitch:union}
    \State \Return $m_{\mathit{sys}}$ \label{alg:stitch:return}
\end{algorithmic}
\caption{\textsc{Stitch}}
\label{fig:stitch}
\end{algorithm}
We assume that, within a set of components $C$, there is a component labeled 
$c_{\mathit{main}}$ that corresponds to the root component in the system 
architectural diagram (e.g., \textit{TC} in our running example).
Algorithm \textsc{Stitch} takes as input $C$, a
set of component-level gFSMs $M$ (one model for each component in
$C$), and a set of logs (one log for each execution)
$L_{\mathit{main}} $ for $c_{\mathit{main}}$; it returns a
system-level gFSM $m_{\mathit{sys}}$.  Internally, \textsc{Stitch} uses
novel auxiliary algorithms (\textsc{Graft}, \textsc{Slice},
\textsc{Insert}), which are described further below.

The algorithm builds a system-level
gFSM $m_\mathit{main}$ for each execution log $l_i \in L_{\mathit{main}}$,
starting from the component-level gFSMs in $M$
(lines~\ref{alg:stitch:beginW}--\ref{alg:stitch:endW}); this is done by  the 
\textsc{Graft} algorithm, described in detail in
\S~\ref{sec:graft}. During the iteration through the execution logs in
$L_{\mathit{main}}$, the resulting system-level
gFSMs $m_\mathit{main}$ are collected in the set $W$.
Last, the gFSMs  in $W$ are merged into $m_{\mathit{sys}}$ using the  
DFA union operation\footnote{One could use the standard DFA minimization 
after the DFA union in line~\ref{alg:stitch:union} to reduce the
size of the system-level gFSM. However, our preliminary evaluation
showed that the minimization operation can reduce the gFSM size (in
terms of numbers of states and transitions) by at most 5\%, and it
increases the execution time of the \textsc{Stitch} algorithm by more
than five times.} (line~\ref{alg:stitch:union}). The algorithm ends
by returning the system-level gFSM  
$m_{\mathit{sys}}$ (line~\ref{alg:stitch:return}), inferred from all
executions in $L_{\mathit{main}}$.

\subsubsection{Graft}
\label{sec:graft}

\begin{figure}
	\centering
	\resizebox{0.9\linewidth}{!}{%
	\begin{tikzpicture}[baseline=0pt,->,>=stealth',shorten >=1pt,auto,node distance=1.2cm, initial text = {}, 
					state/.style={circle, draw, minimum size=0.2mm}]
	\tikzstyle{every state}=[text=black]
	\node[draw] at (-2,1.2) {$m_{c_{X}}$};

 		 \node[state, initial, accepting, initial where=above] at (-2,0) (A) {$s_0$};
 		 \node[state]         (B) [below=0.4cm of A] {$s_1$};
  		\node[state]         (C) [accepting, below =0.4cm of B] {$s_2$};
		\path (A) edge   node {$e_1^X$} (B)
       	 		(B) edge node {$e_2^X$} (C);
		\node[draw] at (-0.25,1.2) {$m_{c_{Y}}$};
		\node[state, initial, accepting, initial where=above] (x)  [right =1cm of A]  {$s_3$};
  		\node[state]         (y) [below =0.4cm of x] {$s_4$};
  		\node[state]         (z) [accepting, below =0.4cm of y] {$s_5$};
		 \path (x) edge    node {$e_1^Y$} (y)
       	 		  (y) edge         node  {$e_3^Y$} (z)
			  edge [loop right] node {$e_2^Y$}();
		\node at (4.2,-1.1) {{Graft}}; 
		\draw [thick, decoration={markings,mark=at position 1 with {\arrow[scale=2,>=stealth]{>}}},postaction={decorate}] (3.6,-1.4) -- (5,-1.4);
		\node[draw] at (8.4,1.2) {$m_{c_{X \upharpoonright Y}}$};
		\node[state, initial, accepting, initial where=above] (a) [right =8cm of x] {$s_0$};
  		\node[state]         (b) [right=1.1cm of a] {$s_1$};
  		\node[state]         (c) [below =0.7cm of b] {$s_4$};
		\node[state]         (d) [left=0.6cm of c] {$s_2$};
  		\node[state]         (e) [accepting, left =0.6cm of d]  {$s_5$};
		
  		\path (a) edge       node {$e_1^X$} (b)
       	 		(b) edge         node  {$e_1^Y$} (c)
			(c) edge         node  {$e_2^X$} (d)
			    edge [loop right] node {$e_2^Y$}()
        			(d) edge              node {$e_3^Y$} (e);
		\node[text width=3.7cm] (T) at (4.2,0.3) 
   			{{Log entries dependencies\\
    				\leavevmode\phantom{ent}$e_1^X \rightsquigarrow \langle e_1^{\mathit{Y}}, e_2^{\mathit{Y}} \rangle$\\
				\vspace{1mm}
  				 \leavevmode \phantom{ent}$e_2^X \rightsquigarrow \langle e_3^Y \rangle$}};	
		\node at (11.2,0.52) {\scriptsize{$\mathit{slice}_1$}}; 
		\node at (7.7,-0.79) {\scriptsize{$\mathit{slice}_2$}}; 
   \begin{pgfonlayer}{background}
   \draw [join=round,black] ($(T.north west) + (-0em, 0em)$) rectangle ($(T.south east) + (0em, 0em)$);
    \draw [join=round,black,dashed] ($(b.north) + (-1.2em, 0em)$) rectangle ($(c.south) + (3.7em, 0em)$);
    \draw [join=round,black,dashed] ($(d.east) + (0.1em, 1.2em)$) rectangle ($(e.west) + (-0.1em, -1.7em)$);
  \end{pgfonlayer}
	\end{tikzpicture}
}
	\caption{The main intuition behind the \textsc{Graft}
          algorithm (for simplicity, we use log entries as transition labels)}
	\label{fig:graft-idea}
\end{figure}
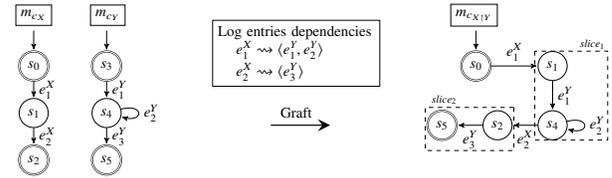

The \textsc{Graft} algorithm builds the system-level gFSM for an
execution by merging the individual component-level gFSMs, taking into
account the log entries dependencies extracted from the execution,
while preserving the component gFSMs as much as possible.
To illustrate the main idea behind the algorithm, let
us consider two components $c_X$ and $c_Y$, whose corresponding gFSMs
(inferred in the pre-processing stage) $m_{c_X}$ and $m_{c_Y}$ are
shown in Figure~\ref{fig:graft-idea}.  These gFSMs respectively accept
log $l_X = \langle e_1^X, e_2^X \rangle$ and log
$l_Y = \langle e_1^Y, e_2^Y, e_3^Y \rangle$. Let us also assume that
in terms of log entries dependencies (expressed through the
\emph{leads-to} relation) we have
$e_1^X \rightsquigarrow \langle e_1^Y, e_2^Y \rangle$ and
$e_2^X \rightsquigarrow \langle e_3^Y \rangle$. Taking into account
these dependencies, intuitively we can say that the gFSM resulting
from the merge of $m_{c_X}$ and $m_{c_Y}$, denoted by
$m_{c_{X \upharpoonright Y}}$, should accept the sequence of log
entries $\langle e_1^X, e_1^Y, e_2^Y, e_2^X, e_3^Y\rangle$. To obtain
$m_{c_{X \upharpoonright Y}}$, we first ``slice'' $m_{c_Y}$ into
two gFSMs: $\mathit{slice}_1$ (accepting
$\langle e_1^Y, e_2^Y \rangle$) and $\mathit{slice}_2$ (accepting
$\langle e_3^Y \rangle$); then, we ``insert'' 1) $\mathit{slice}_1$
as the target of the transition of $m_{c_X}$ that reads $e_1^X$, and 2)
$\mathit{slice}_2$ as the target of the transition of $m_{c_X}$ that reads
$e_2^X$. Note that the self-loop transition in $m_{c_Y}$ is preserved
in $m_{c_{X \upharpoonright Y}}$ as a result.

As shown in Algorithm~\ref{fig:graft},  \textsc{Graft}  takes as input a component $c_{\mathit{cur}}$, an execution log 
$l_{\mathit{cur}} = \langle e_1, \dots, e_z \rangle$, and a set of 
component-level gFSMs $M = \{m_{c_\mathit{main}}, m_{c_1}, \dots, m_{c_n}\}$; it 
returns a gFSM $m_{\mathit{sl}}$ that accepts the sequence of log entries 
composed of the entries $e_i \in l_{\mathit{cur}}$, with each $e_i$ interleaved 
with the log entries to which it \emph{leads-to}.

\begin{algorithm}[t]
\footnotesize
\begin{algorithmic}[1]
\Input{Component $c_{\mathit{cur}}$ \newline
\phantom{Co} Log $l_{\mathit{cur}} = \langle e_1, \dots, e_z
\rangle$ \newline
\phantom{Co} Set of gFSMs $M = \{m_{c_{\mathit{main}}}, m_{c_1}, \dots, m_{c_n}\}$}
\Output{System model for the current execution  $m_{\mathit{sl}}$}

    \State gFSM $m_{\mathit{cur}} \gets \mathit{getComponentGFSM}(M,c_{\mathit{cur}})$
    \State gFSM $m_{sl} \gets$ \textsc{Slice}($m_{c_\mathit{cur}}, l_{\mathit{cur}}$)\label{alg:graft:slice}
    \State State $s \gets \mathit{getInitialState}(m_{sl})$\label{alg:graft:initState}
    \ForEach{$e_i \in l_{\mathit{cur}}$}\label{alg:graft:begini}
        \State GuardedTransition $\mathit{gt} \gets \mathit{getGuardedTran}(m_{sl}, s, e_i)$\label{alg:graft:gt}
            \State Set of gFSMs $W \gets \emptyset$
            \ForEach{log entries sequence $l_d \mid e_i \rightsquigarrow l_d$}\label{alg:graft:beginld}
            \State Component $c_d \gets \mathit{getComponentFromLog}(l_d)$
            \State gFSM $m_g \gets$ \textsc{Graft}($c_d, l_d, M$)\label{alg:graft:rec}
                \State $W \gets  \{m_g\} \cup W$ \label{alg:graft:addW}
            \EndFor
            \State gFSM $m_{\mathit{pl}} \gets \mathit{DFAParallelComposition}(W)$\label{alg:graft:pcomp}
            \State $m_{sl} \gets$ \textsc{Insert}($m_{sl}, \mathit{gt}, m_{pl}$)\label{alg:graft:insert}
        \State $s \gets \mathit{getTargetState}(\mathit{gt})$\label{alg:graft:nexts}
    \EndFor
    \State \textbf{return} $m_{\mathit{sl}}$
\end{algorithmic}
\caption{Graft}
\label{fig:graft}
\end{algorithm}

The algorithm starts by slicing the gFSM $m_{c_{\mathit{cur}}}$ of the
input component $c_{\mathit{cur}}$ into a gFSM $m_{\mathit{sl}}$  that accepts
only $l_{\mathit{cur}}$ (line~\ref{alg:graft:slice}); the actual slicing is done through algorithm
\textsc{Slice}, described in detail in \S~\ref{sec:slice}. The rest
of the algorithm expands $m_{\mathit{sl}}$ taking into account the
log entries dependencies (lines~\ref{alg:graft:initState}--\ref{alg:graft:nexts}): 
for each log entry $e_i \in l_{\mathit{cur}}$, a gFSM $m_g$
that accepts the log entries sequence that $e_i$ \emph{leads-to} is built and 
``inserted'' in $m_{\mathit{sl}}$ as the target of the guarded transition 
$\mathit{gt}$ that reads $e_i$. More precisely, the algorithm performs a run of
$m_{\mathit{sl}}$ as if it were to accept the log $l_{\mathit{cur}}$: starting
from the initial state of $m_{sl}$ (line~\ref{alg:graft:initState}),
it moves to the next state $s$ by making the guarded transition
$\mathit{gt}$ that reads $e_i$ (line~\ref{alg:graft:gt}). As part of this
move, for each log entry sequence $l_d$ such that
$e_i \rightsquigarrow l_d$, we recursively call \textsc{Graft} to
build the gFSM $m_g$ that accepts $l_d$; this gFSM is then added to
the set $W$ (lines~\ref{alg:graft:beginld}--\ref{alg:graft:addW}) .
Since a log entry $e_i$ may \emph{lead-to} log entries sequences of
multiple components, we compose the individual gFSMs in $W$ using the
standard DFA parallel composition operation
(line~\ref{alg:graft:pcomp}). The resulting gFSM $m_{\mathit{pl}}$ is
``inserted'' in $m_{\mathit{sl}}$ as the target of $\mathit{gt}$
by the \textsc{Insert} algorithm (line~\ref{alg:graft:insert}),
described in detail in \S~\ref{sec:insert}. At the end of each
iteration of the loop, the state $s$ is updated with the target state
of the $\mathit{gt}$ transition (line~\ref{alg:graft:nexts}).

\begin{figure}
	\centering
	\resizebox{\linewidth}{!}{\input{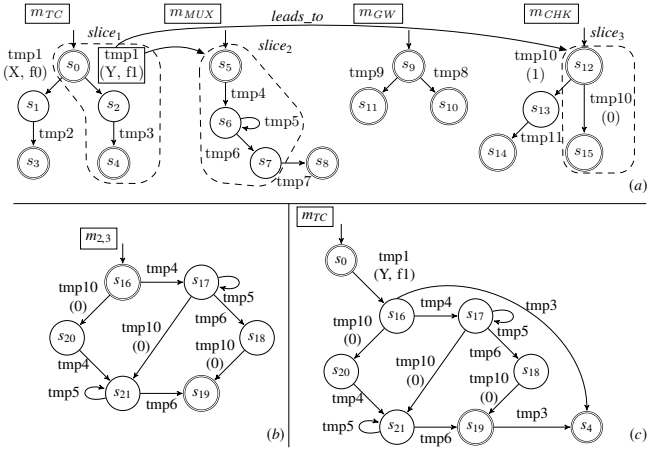}}
		\caption{Application of algorithm \textsc{Graft} to
          Execution-2 of the running example}
	\label{fig:graft-example}
\end{figure}

As an example, let us consider the case in which the \textsc{Stitch}
algorithm calls the \textsc{Graft} algorithm when processing
Execution-2 of our running example.
Figure~\ref{fig:graft-example}-(a) shows the component-level gFSM and
how they are related when taking into account the \emph{leads-to}
relation listed in Table~\ref{tab:causality}. Algorithm
\textsc{Stitch} invokes \textsc{Graft} with parameters
$c_{\mathit{cur}} = \mathit{TC}$,
$l_{\mathit{cur}} = \langle e_{1,2}^{\mathit{TC}},
e_{2,2}^{\mathit{TC}} \rangle$,
$M = \{m_{\mathit{TC}}, m_{\mathit{MUX}}, m_{\mathit{CHK}},
m_{\mathit{GW}}\}$. 
The call to \textsc{Slice} yields the gFSM $\mathit{slice_1}$ shown in 
Figure~\ref{fig:graft-example}-(a); it accepts 
$\langle e_{1,2}^{\mathit{TC}}, 
e_{2,2}^{\mathit{TC}} \rangle$, using the transitions labeled with
$\mathit{tmp}_1(Y,f1)$ and $\mathit{tmp}_3$. 
Then, starting from $s_0$ of $\mathit{slice}_1$,  the invocation of the auxiliary function
\emph{getGuardedTran} yields the guarded transition $(s_0, \mathit{tmp}_1, [Y, f1], s_2)$ that reads 
$e_{1,2}^{\mathit{TC}}$. Since $e_{1,2}^{\mathit{TC}} \rightsquigarrow 
\langle e_{1,2}^{\mathit{MUX}}, e_{2,2}^{\mathit{MUX}}, e_{3,2}^{\mathit{MUX}} \rangle$ and $e_{1,2}^{\mathit{TC}} \rightsquigarrow 
e_{1,2}^{\mathit{CHK}}$, the algorithm makes a recursive call for 
$\langle e_{1,2}^{\mathit{MUX}}, e_{2,2}^{\mathit{MUX}}, e_{3,2}^{\mathit{MUX}} \rangle$, 
which returns the sliced gFSM 
$\mathit{slice}_2$, and for $\langle e_{1,2}^{\mathit{CHK}} \rangle$, which 
returns $\mathit{slice}_3$; both gFSMs are shown in 
Figure~\ref{fig:graft-example}-(a). 
At the end of the inner loop, we
have $W=\{\mathit{slice}_2,\mathit{slice}_3\}$; their parallel
composition is $m_{2,3}$ and is shown in
Figure~\ref{fig:graft-example}-(b). 
This gFSM is then inserted in $\mathit{slice}_1$ as the target of the 
transition $(s_0, \mathit{tmp}_1, [Y, f1], s_2)$, as shown in 
Figure~\ref{fig:graft-example}-(c).
The algorithm ends for $e_{1,2}^{\mathit{TC}}$ by inserting $m_{2,3}$ in $s_2$ and 
moves on to the next log entry $e_{2,2}^{\mathit{TC}}$.

\subsubsection{Slice}\label{sec:slice}
This algorithm takes as input a component-level gFSM 
$m_c$ and a log $l_c$; it returns a new gFSM $m_{\mathit{sl}}$, which
is the sliced version of $m_c$ and accepts only $l_c$. 

Its pseudocode is shown in Algorithm~\ref{fig:slice}. First, the
algorithm retrieves the state of $m_c$ that will become the 
initial state $s$ of the sliced gFSM $m_{\mathit{sl}}$ 
(line~\ref{alg:slice:init}). Upon the first invocation of \textsc{Slice} for a
certain gFSM $m_c$, $s$ will be the initial state of $m_c$; for
the subsequent invocations, $s$ will be the last state visited in $m_c$ when running 
the previous slice operations. Starting from $s$, the algorithm
performs a run of $m_c$ as if it were to accept the log $l_c$: the
traversed states and guarded transitions of $m_c$ are added into
$m_{\mathit{sl}}$ (lines~\ref{alg:slice:beginfor}--\ref{alg:slice:endfor}).
At the end of the loop, the algorithm records
(line~\ref{alg:slice:update}) the last state visited in $m_c$ when
doing the slicing, which will be used as the initial state of the next
slice on $m_c$; it then ends by returning $m_{\mathit{sl}}$.  

\begin{algorithm}[t]
  \footnotesize
\begin{algorithmic}[1]
\Input{A component gFSM $m_c$ \newline
\phantom{aai} A component Log $l_c = \langle e_1, \dots, e_z\rangle$}
\Output{a sliced gFSM $m_{\mathit{sl}}$}

    \State gFSM $m_{\mathit{sl}} \gets \mathit{initGFSM}()$
    \State State $s \gets \mathit{getSliceStartState}(m_c)$\label{alg:slice:init}
    \ForEach{$e_i \in l_c$}\label{alg:slice:beginfor}
        \State Guarded Transition $\mathit{gt} \gets \mathit{getGuardedTran}(m_c, s, e_i)$
        \State $m_{\mathit{sl}} \gets \mathit{AddGuardedTranAndStates}(m_{\mathit{sl}}, \mathit{gt})$
        \State $s \gets \mathit{getTargetState}(\mathit{gt})$\label{alg:slice:endfor}
    \EndFor
    \State $ \mathit{updateSliceStartState}(m_c,s)$\label{alg:slice:update}
    \State \textbf{return} $m_{\mathit{sl}}$\label{alg:slice:return}
\end{algorithmic}
\caption{Slice}
\label{fig:slice}
\end{algorithm}

\subsubsection{Insert}
\label{sec:insert}
We recall that this algorithm is invoked by the \textsc{Graft}
algorithm to ``insert'' a gFSM $m_y$ into a gFSM $m_x$ as the 
target of a guarded transition $\mathit{gt}$ of $m_x$, taking
into account the log entries dependencies. More specifically, let us
consider a log entry $e$ and a set of logs $L = \{l_1, \dots, l_n\}$
where $e \rightsquigarrow l_i$ for $i=1,\dots,n$; the transition
$\mathit{gt}$ of $m_x$ reads $e$, and $m_y$ is the parallel
composition of the gFSMs that accepts the logs in $L$.  The
\textsc{Insert} algorithm merges $m_y$ into $m_x$ such that, by
``inserting'' $m_y$ as the target of the guarded transition $\mathit{gt}$, $m_x$
can read the (entries in the) logs in $L$ right after reading $e$.

\begin{figure}
	\centering
	\usetikzlibrary{arrows,positioning,shapes,backgrounds,fit}  

\tikzset{mycircled/.style={circle,draw,inner sep=0.1em,line width=0.04em}}

\resizebox{0.8\linewidth}{!}{%
	\begin{tikzpicture}[baseline=0pt,->,>=stealth',shorten >=1pt,auto,node distance=1.1cm, initial text = {}, 
					state/.style={circle, draw, minimum size=0.2mm}]
	\tikzstyle{every state}=[text=black]
	\node[draw] at (-1.75,1) {$m_x$};
 		 \node[state] at (-1.3, 0) (A) {$s_p$};
 		 \node[state]         (B) [below of=A] {$s_t$};
  		\node[state]         (C) [below of=B] {$s_n$};
		\path (A) edge   node {$a$} (B)
       	 		(B) edge node {$b$} (C);
		\draw[->, bend left=35]($(C.west) + (0em, 0em)$) to node[right] {$d$} ($(A.west) + (0em, 0em)$);
		\draw [->, dotted] (-1.3,0.8) -- (-1.3,0.35); 
		\draw[dotted,shorten >=3pt]  (C.south) -- ++(0,-0.5cm);
	
	\node[draw] at (-0.45,0.5) {$m_y$};
 	\node[state] (a1) at (0, -0.5) {$s_i$};
 	 \node[state]         (b1) [accepting, below of=a1] {$s_f$};
	\path (a1) edge   node [left] {$\alpha$} (b1);
	\draw [->, dotted] (0,0.3) -- (0,-0.18); 

	\node at (0.9,-0.9) {1};
	\draw [thick, decoration={markings,mark=at position 1 with {\arrow[scale=2,>=stealth]{>}}},postaction={decorate}] (0.6,-1.1) -- (1.4,-1.1);
	
	\node[draw] at (2,1) {$m_{x}$};
 	 \node[state] (x) [right =3cm of A] {$s_p$};
 	 \node[state]         (y) [below of=x] {$s_t$};
  	\node[state]         (z) [below of=y] {$s_n$};
	\node[state] (a2) at (3.5, -0.5) {$s_i$};
 	 \node[state]         (b2) [below of=a2] {$s_f$};
	\path (x) edge   node {$a$} (y)
		(y) edge   node {$b$} (z)
		(x) edge   node {$a$} (a2)
       	 	(a2) edge node {$\alpha$} (b2)
		(b2) edge node {$b$} (z);
	\draw[->, bend left=35]($(z.west) + (0em, 0em)$) to node[right] {$d$} ($(x.west) + (0em, 0em)$);
	\draw[dotted,shorten >=3pt]  (z.south) -- ++(0,-0.5cm);
	\draw [->, dotted] (2.45,0.8) -- (2.45,0.35); 
	
	\node at (4.4,-0.9) {2};
	\draw [thick, decoration={markings,mark=at position 1 with {\arrow[scale=2,>=stealth]{>}}},postaction={decorate}] (4.1,-1.1) -- (4.9,-1.1);
	\node[draw] at (5.5,1) {$m_x$};
 	 \node[state] (x) [right =6.5cm of A] {$s_p$};
 	 \node[state]         (y) [below of=x] {$s_m$};
  	\node[state]         (z) [below of=y] {$s_n$};
 	 \node[state]         (b2) at (7.1, -1.4) {$s_f$};
	\path (x) edge   node {$a$} (y)
		(y) edge   node {$b$} (z)
       	 	(y) edge node [above]{$\alpha$} (b2)
		(b2) edge node {$b$} (z);
	\draw[->, bend left=35]($(z.west) + (0em, 0em)$) to node[right] {$d$} ($(x.west) + (0em, 0em)$);
	\draw[dotted,shorten >=3pt]  (z.south) -- ++(0,-0.5cm);
  	\draw [->, dotted] (5.95,0.8) -- (5.95,0.35); 
	\end{tikzpicture}}
	\caption{Example showing the basic idea of the \textsc{Insert}
          algorithm, when inserting $m_y$ into $m_x$ as the target of
          the guarded transition $\mathit{gt}$ with
          $\mathit{gt} = (s_p, a, s_t)$. Step~1 shows the application
          of duplication and redirection; step~2 applies
          determinization to merge states $s_t$ and $s_i$. }
	\label{fig:insert-example}
\end{figure}
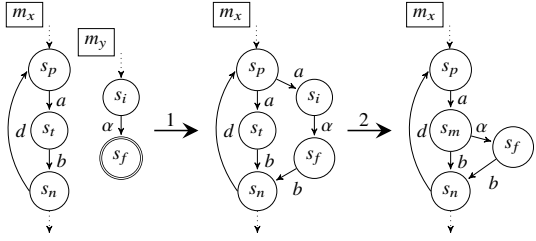

We illustrate how the algorithm works through the example in
Figure~\ref{fig:insert-example}, in which the input gFSMs $m_x$ and
$m_y$ are shown on the left side; we will insert $m_y$ into $m_x$ as
the target of the guarded transition $\mathit{gt}$, labeled with $a$
and having $s_t$ as target state. Without loss of generality, we
assume that $m_y$ has only one transition (labeled with $\alpha$)
between its initial state $s_i$ and the final one $s_f$.  The main
idea behind the \textsc{Insert} algorithm is to duplicate both
incoming and outgoing transitions of the target state of
$\mathit{gt}$, and to redirect the new copies to the initial and
finals states of $m_y$. More specifically:
\begin{compactitem}
\item the incoming transition $\mathit{gt}$ of $s_t$ (labeled with $a$) is duplicated 
and the new copy is redirected, by changing its target state, to the initial state of $m_y$ 
(i.e., $s_i$);
\item the outgoing transitions of $s_t$ (e.g., the one labeled with 
$b$) are duplicated and the new copies are redirected, by changing the source state, such
that they originate from the final state of $m_Y$ (i.e., $s_f$).
\end{compactitem}
The updated $m_x$, resulting from the application of duplication and
redirection, is shown in the middle of
Figure~\ref{fig:insert-example}.  We remark that we keep the original
incoming and outgoing transitions of $s_t$ on purpose, to take into
account the cases in which one of the log entries read by
$\mathit{gt}$ does not \emph{lead-to} log entries read by the
transition labeled with $\alpha$.  Duplication and redirection
operations introduce some nondeterminism in $m_x$; in our example, $s_p$ has two
outgoing transitions both labeled with $a$. We remove nondeterminism
using a \emph{determinization} procedure~\cite{1566607}, which
recursively merges pair of states that introduces
nondeterminism\footnote{This procedure is different from the standard
  NFA  (non-deterministic finite automaton) to DFA conversion since it
  yields an automaton which may accept a more general language than
  the NFA it starts from~\cite{1566607}.}; in our example, the
determinization procedure will merge $s_t$ and $s_i$.  The final $m_x$
is shown on the right side of Figure~\ref{fig:insert-example}.

Algorithm~\ref{fig:insert} shows the pseudocode of the \textsc{Insert}
algorithm.  The algorithm takes a gFSM $m_x$, a guarded transition
$\mathit{gt}$, and a gFSM $m_y$; it returns the updated $m_x$ that
includes $m_y$ as the target of $\mathit{gt}$.  In the algorithm,
$s_t$ is the target state of $\mathit{gt}$, $s_i$ is the initial state
of $m_y$ and $F_y$ is the set of the final states of $m_y$. The core
part (lines~\ref{alg:insert:beginredir}--\ref{alg:insert:endredir})
iterates through each guarded transition $t$ of $s_t$, duplicates it,
and redirects the new copy as described above, using the the auxiliary
function $\mathit{duplicateAndRedirectTransitions}$.  Last, the
algorithm removes nondeterminism using $\mathit{determinize}$
(line~\ref{alg:deter}); it ends by returning the updated gFSM $m_x$
(line~\ref{alg:insert:return}).

\begin{algorithm}[t]
\footnotesize
\begin{algorithmic}[1]
  \Input{gFSM $m_x$ \newline
  \phantom{iiii}  Guarded Transition $\mathit{gt}$ \newline
 \phantom{iiii} gFSM $m_y$}
\Output{Updated gFSM $m_x$}

    \State State $s_t \gets\mathit{getTargetState}(\mathit{gt})$\label{alg:insert:initStart}
    \State State $s_i \gets \mathit{getInitialState}(m_y)$
    \State Set of States $F_y \gets \mathit{getFinalStates}(m_y)$\label{alg:insert:initEnd}    

    \ForEach{Guarded Transition $t$ of $s_t$}\label{alg:insert:beginredir}
    		\If{$t = \mathit{gt}$}
        		\State $\mathit{duplicateAndRedirectTransitions}(t, s_t, \{s_i\})$
        	\ElsIf{$t$ is an outgoing transition}
            		\State $\mathit{duplicateAndRedirectTransitions}(t, s_x, F_y)$
        \EndIf
    \EndFor\label{alg:insert:endredir}
    \State $\mathit{determinization}(m_x)$\label{alg:deter}
    \State \textbf{return} $m_x$\label{alg:insert:return}
\end{algorithmic}
\caption{Insert}
\label{fig:insert}
\end{algorithm}

\paragraph*{Accuracy of the system-level gFSM}
\app has three main sources of over-generalization that reduce
the accuracy:
\begin{inparaenum}[(1)]
\item  component-level model inference,
\item parallel composition in \textsc{Graft}, and
\item determinization in \textsc{Insert}.
\end{inparaenum}
The first is essentially inevitable in any model inference algorithm;
we try to compensate it by using a state-of-the-art tool (MINT) to
infer component models that are as accurate as possible.  The second
source may become a problem when the log dependencies identified in
the preprocessing stage are incorrect; nevertheless,
over-generalization caused by parallel composition is limited because
the latter is only performed on the sliced gFSMs.  The last source has
limited effects because recursive determinization rarely occurs in
practice.

We further discuss the accuracy of \app corroborated by
experimental data in the next section.

\section{Evaluation}
\label{sec:evaluation}
We have implemented the \app approach as a Python program. In this
section, we report on the evaluation of the performance of the \app
implementation in generating the model of a component-based system
from the individual component-level logs.

First, we assess the scalability of \app  
in inferring models from large execution logs. This is the primary 
dimension we focus on since we propose \app as a viable alternative to
state-of-the-art techniques for processing large logs. 
Second, we analyze how accurate the models generated by \app are. 
This is an important aspect because it is orthogonal to scalability
and has direct implications on the possibility of using the models
generated by \app in other software engineering tasks (e.g., test case generation).
Summing up, we investigate the following research questions:
\begin{compactenum}[RQ1:]
\item \textit{How scalable is \app when compared to state-of-the-art model
    inference techniques?}
\item \textit{How accurate are the models (in the form of gFSMs)
    generated by \app when compared to those generated by state-of-the-art
    model inference techniques?}
\end{compactenum}

\subsection{Benchmark and Evaluation Settings}
We used a benchmark composed of industrial, proprietary datasets
provided by one of our industrial partners, active in the satellite
industry. The benchmark contains component-level logs recorded during
the execution of a satellite ground control system, which includes
six major components.
We created the benchmark as follows. First, we 
executed system-level tests on the ground control system 120 times and,
in each test execution, we collected the log files of the six major components.
Then, we created seven datasets of size ranging from 5K to 35K, where the size
is expressed in terms of the total number of log entries. We assembled
each dataset by randomly selecting a number of executions out of the
pool of 120 executions, such that the total size of the logs contained
in the dataset matched the desired dataset size. By construction, each dataset contains
logs of the six major components of the system.
The first three columns of Table~\ref{tab:scalable} show, 
for each dataset in our benchmark, the size and the number of executions included in it.
In total, there are 92 unique templates (i.e., unique number of events) for all logs.
All the collected logs (anonymized) as well as the evaluation results are available at 
\url{http://tiny.cc/SANER20-SCALER}.
The experiments have been executed on a high-performance computing
platform, using one of its quad-core nodes running CentOS 7 
on a \SI{2.4}{\giga\Hz} Intel Xeon E5-2680 v4 processor with \SI{4}{\giga\byte} memory.

\subsection{Scalability}
\label{sec:scale}

\subsubsection{Methodology}

To answer RQ1, we assess the scalability of \app, in terms of
execution time with respect to the size of the logs, in comparison with
MINT~\cite{walkinshaw2016inferring}, a state-of-the-art model
inference tool. We selected MINT as baseline because other tools
are either not publicly available or require information not
available in most practical contexts, including ours (e.g., channels' definitions; see 
section~\ref{sec:example}).

We ran both tools to infer a system-level model for each dataset in
our benchmark. We provided as input to \app 1) the logs of the six
components recorded in the executions contained in each dataset;
2) the architectural dependencies among components; 3) the list of
log message templates for communication events, received from a domain
expert. As for MINT, we
provided as input the system-level logs of the system executions contained in
each dataset. We derived these system-level logs by linearizing the
individual component logs in each execution, taking into account the
log entries dependencies. To guarantee a fair comparison, these
dependencies are the same as those extracted in the pre-processing
stage of \app.  Since the total number of possible system-level logs
is extremely large due to the linearization of the parallel behaviors
of the components, we only considered one system-level log for each
execution.

We remark that we used \emph{two} instances of MINT: the one used
internally by \app to generate component-level models; the other one
for the comparison in inferring system-level models. For both
instances, we used the default configuration (i.e., state merging
threshold $k=2$ and J48 as data classifier algorithm)~\cite{walkinshaw2016inferring}.
Furthermore, to identify the
event templates required by the MINT instances to parse the log
entries, we first used a state-of-the-art tool
(MoLFI~\cite{messaoudi2018search}) to compute them and then we asked a
domain expert to further refine them, e.g., by collapsing similar
templates into a single one.
To take into account the randomness of the log linearization (i.e., 
only one linearized system-level log) for each execution of MINT, we 
ran both MINT and \app ten times on each dataset.
For each run, we set an overall time out of 24h for the model inference 
process both for MINT and for \app.

To assess the statistical significance of the difference between the execution 
time of \app and MINT (if any), we used the non-parametric Wilcoxon rank 
sum test with a level of significance $\alpha=0.05$. 
Furthermore, we used the Vargha-Delaney ($\hat{A}_{12}$) statistic for
determining the effect size of the difference.  In our case,
$\hat{A}_{12}<0.5$ indicates that the execution time of \app is lower
than that of MINT.

\subsubsection{Results}
The columns under the header ``Scalability'' of
Table~\ref{tab:scalable} show the scalability results for \app and
MINT. More precisely, column \emph{MINT} indicates the 
execution time of MINT;
columns \emph{Prep}, \emph{Stitch}, and
\emph{Total} indicate the average (over the ten runs) execution
time (in seconds) and the corresponding standard deviation of 
\app for the pre-processing stage, the stitching stage, and the
cumulative execution time, respectively; column \emph{SpeedUp} reports the 
speedup of \app over MINT computed as
$
\tfrac{\mathit{Time}_{\mathit{MINT}}}{\mathit{Time}_{\mathit{\app}}}
$~\cite{sahni1996performance}.

{\renewcommand{\arraystretch}{1.4} 
\begin{table*}
 \centering
    \caption{Execution time (in seconds), recall, and specificity of \app and MINT}
\label{tab:scalable}
 \resizebox{\linewidth}{!}{%
    \setlength\tabcolsep{3.5mm}
    \begin{filecontents*}{results/new_expr_results.csv}
Dataset,Size,no_ex,mint_states,scaler_states,ratio,mint_trans,scaler_trans,ratio,scmint,scprep,scstitch,sctotal,speedup,rcmint,rcscaler,rcdelta,spmint,spscaler,spdelta
D05K,5058,13,617.7,3816,6.2,837.2,7295,8.7,319.6,6.0,5.8,11.8,27.1,0.09,0.65,56,1.00,0.98,-2
D10K,10208,28,1207.3,6647,5.5,1585.4,12720,8.0,2597.0,10.9,15.3,26.2,99.2,0.16,0.63,47,0.99,0.98,-1
D15K,15078,42,1582.6,4184,2.6,2064.4,8868,4.3,7403.8,13.4,19.8,33.2,222.8,0.52,0.82,30,0.99,0.97,-2
D20K,20094,56,2257.0,7463,3.3,2914.7,15851,5.4,16022.2,18.6,32.1,50.7,315.9,0.58,0.86,28,0.98,0.97,-1
D25K,25034,71,3067.2,9496,3.1,3976.3,18767,4.7,35378.6,24.3,58.2,82.5,428.7,0.56,0.83,27,0.98,0.96,-2
D30K,30103,86,2871.3,19467,6.8,3701.8,40204,10.9,59222.7,29.0,129.6,158.6,373.3,0.61,0.86,25,0.98,0.96,-2
D35K,35079,101,,10432,,,19707,,timeout,32.5,72,104.5,,,0.88,,,0.97,
Avg,20093.4,56.7,1933.9,8786.4,4.6,2513.3,17630.3,7.0,20157.3,19.3,47.5,66.8,244.5,0.42,0.79,35.5,0.99,0.97,-1.67
\end{filecontents*}
\pgfplotstabletypeset[
    skip rows between index={0}{1},
     font={\scriptsize},
      empty cells with={N/A},
    	header=false,
        col sep=comma,
        every head row/.style={
        output empty row, after row={\midrule},
        before row={
        \toprule
\multirow{3}{*}{Dataset} & \multirow{3}{*}{Size} & \multirow{3}{*}{Exec} & \multicolumn{6}{c}{System-level gFSM} & \multicolumn{5}{c}{Scalability} & \multicolumn{6}{c}{Accuracy} \\
		\cmidrule(rr){4-9}\cmidrule(rr){10-14}\cmidrule(rr){15-20}
 & & & \multicolumn{3}{c}{States} & \multicolumn{3}{c}{Transitions} & \multirow{2}{*}{MINT} & \multicolumn{3}{c}{\app} & \multirow{2}{*}{SpeedUp} &  \multicolumn{3}{c}{Recall} & \multicolumn{3}{c}{Specificity} \\
		\cmidrule(rr){4-6}\cmidrule(rr){7-9}\cmidrule(rr){11-13}\cmidrule(rr){15-17}\cmidrule(rr){18-20}
	 &  &  & 
	 MINT & \app & Ratio & MINT & \app & Ratio & &
	 Prep(\si{\s}) & Stitch(\si{\s}) & Total(\si{\s}) & & 
	 MINT &  \app & $\Delta_{\text{R}}$($pp$) & MINT &  \app & $\Delta_{\text{S}}$($pp$)\\
             }
    },
   columns={0,1,2,3,4,5,6,7,8,9,10,11,12,13,14,15,16,17,18,19},
   display  columns/{0}/.style={column name={},string type, column type={c@{\hskip 1mm}},},
    display columns/1/.style={column name={},string type, column type={r@{\hskip 1mm}}},
    display columns/2/.style={column name={},string type,column type={r@{\hskip 3mm}},},
    display columns/3/.style={column name={},string type, column type={r@{\hskip 1mm}},},     
    display columns/4/.style={column name={},string type, column type={r@{\hskip 1mm}},},     
    display columns/5/.style={column name={},string type, column type={r@{\hskip 3mm}},},     
    display columns/6/.style={column name={},string type, column type={r@{\hskip 1mm}},},     
    display columns/7/.style={column name={},string type, column type={r@{\hskip 1mm}},},     
    display columns/8/.style={column name={},string type, column type={r@{\hskip 3mm}},},      
    display columns/9/.style={column name={},string type, column type={r@{\hskip 1mm}},},     
    display columns/10/.style={column name={},string type, column type={r@{\hskip 1mm}},},
    display columns/11/.style={column name={}, string type, column type={r@{\hskip 1mm}},},
    display columns/12/.style={column name={}, string type, column type={r@{\hskip 1mm}},},
    display columns/13/.style={column name={}, string type, column type={r@{\hskip 3mm}},},
    display columns/14/.style={column name={}, string type, column type={r@{\hskip 1mm}},},     
    display columns/15/.style={column name={}, string type, column type={r@{\hskip 1mm}},},     
    display columns/16/.style={column name={}, string type, column type={r@{\hskip 3mm}},},     
    display columns/17/.style={column name={}, string type, column type={r@{\hskip 1mm}},},     
    display columns/18/.style={column name={}, string type, column type={r@{\hskip 1mm}},},     
    display columns/19/.style={column name={}, string type, column type={r@{\hskip 1mm}},},
         every last row/.style={after row=\bottomrule}, 
         every row no 7/.style={before row=\midrule},
    ]{results/new_expr_results.csv}}
\end{table*}

\app is faster than MINT for all the datasets in our benchmark; 
the speed-up ranges between 27x (for the dataset D05K) and 428x 
(for the dataset D25K). The speed-up increases with the size of the datasets and, 
thus, the benefit of using \app over MINT increases for larger logs. 
Note that MINT reached the time out for the largest dataset (D35K) 
without producing any model. The Wilcoxon test also 
confirms that the differences in execution time between \app and MINT are 
statistically significant ($p$-value $<0.01$ for all datasets) 
and the Vargha-Delaney statistic indicates that the effect size is always 
\textit{large} ($\hat{A}_{12}<0.10$) for all datasets.

Analyzing the performance of the two instances of MINT, we can say
that when MINT is used for component-level model inference is much
faster than MINT used for system-level model inference because (1) the
component logs are smaller than the system-level logs and (2) there is
a higher similarity among component logs than system-level logs.

\subsection{Accuracy}
\label{sec:accur}

\subsubsection{Methodology}
To answer RQ2, we ran both MINT and \app 
to evaluate and compare their accuracy for each dataset, 
in terms of recall and specificity of the inferred models following previous
studies~\cite{walkinshaw2016inferring, mariani2017gk,Emam:2018:IEP:3208361.3196883}.  
Recall measures the ability of the inferred models of a system to accept ``positive''
logs; specificity measures the ability of the inferred models to reject ``negative'' logs.
We computed these metrics by using the well-known $k$-folds cross
validation method, which has also been used in previous
work~\cite{walkinshaw2016inferring,
  mariani2017gk,Emam:2018:IEP:3208361.3196883} in the area of model
inference. This method randomly partitions a set of logs into $k$
non-overlapping folds: $k-1$ folds are used as input of the model
inference tool, while the remaining fold is used as ``test set'', to
check whether the model inferred by the tool accepts the logs in the
fold. The procedure is repeated $k$ times until all folds have been
considered exactly once as the test set. For each fold, if the inferred
model successfully accepts a positive log in the test set, the positive
log is classified as True Positive (TP); otherwise, the positive log
is classified as False Negative (FN). Similarly, if an inferred model
successfully rejects a negative log in the test set, the negative log is
classified as True Negative (TN); otherwise, the negative log is
classified as False Positive (FP).  Based on the classification
results, we calculated the recall (R) as
$\mathit{R}=\tfrac{|\mathit{TP}|}{|\mathit{TP}|+|\mathit{FN}|}$, 
and the specificity (S) as
$\mathit{S=\tfrac{|\mathit{TN}|}{|\mathit{TN}|+|\mathit{FP}|}}$.

As done in previous work~\cite{walkinshaw2016inferring,
  mariani2017gk,Emam:2018:IEP:3208361.3196883}, we synthesized
negative logs from positive logs by introducing small changes
(mutations): 1) swapping two randomly selected log entries, 2)
deleting a randomly selected log entry, and 3) adding a log entry
randomly selected from other executions. To make sure a log resulting
from a mutation contains invalid behaviors of the system, we checked
whether the sequence of entries around the mutation location (i.e.,
the mutated entries and the entries immediately before and after the
mutants) did not also appear in the positive logs.

Note that we needed to derive system-level logs from the individual 
component logs in test sets to check the acceptance of the system-level 
models inferred by \app and MINT. To this end, as done for the 
scalability evaluation, for each execution in the test sets, we linearized 
the individual component logs to derive the system-level log. Also, to 
take into account the randomness of the derivation of system-level 
logs, we repeat the 10-folds cross validation ten times on each dataset 
and then applied statistical tests as done for the scalability 
evaluation.

\subsubsection{Results}
The columns under the header ``Accuracy'' of Table~\ref{tab:scalable} show 
the results of MINT and \app in terms of recall, specificity, and
difference of these values (in percentage points, $pp$) between \app and MINT. 

MINT achieves high specificity scores, always greater than $0.98$. 
However, recall is low, ranging between 0.09 for the D05K dataset 
and 0.61 for the D30K dataset. 
Notice that no results were obtained for the larger dataset with 35K log entries
because MINT reached the timeout of 24h without generating any model.
\app achieves a slightly lower specificity than MINT, 
with an average difference of 1.67$pp$.
However, \app achieves substantially higher recall than MINT. 
The difference in recall values ranges between +25$pp$ (D30K dataset) 
and +56$pp$ (D05K dataset), with an average improvement of 35.5$pp$.
Such a result can be explained mainly because MINT takes as input only one system-level log 
among all possible instances of the linearization of the parallel behaviors of the components
for each execution, and fails to scale up to take as input all the possible
system-level logs.
Related to this, since \app takes as input all the possible system-level logs
(in the form of component-level logs with the log entries dependencies)
for each execution, it returns as output a system-level gFSM having on average
4.6x more states and 7.0x more transitions than MINT
(see the columns under the header 
``System-level gFSM'' in Table~\ref{tab:scalable}).

According to the Wilcoxon test, \app always achieves a statistically 
higher recall than MINT for all datasets ($p\text{-value}<0.01$) with a 
large effect size. However, \app achieves a statistically lower 
specificity than MINT in five out of seven datasets (i.e., with 5K, 
10K, 20K and 30K log entries). While the 
difference in specificity are statistically significant, it is worth 
noting that the magnitude of the difference is small, being no larger 
than 2$pp$.

\subsection{Discussion and Threats to Validity}
\label{sec:eval-discuss}
From the results above, we conclude that, for the large logs typically 
encountered in practice, SCALER provides results that are good enough 
to generate nearly correct (with a specificity always greater than 0.96) 
and largely complete models (with an average recall of 0.79).

The incompleteness of the inferred models is due to the
limited knowledge we have on the system (i.e., the incomplete list of
message templates characterizing communication events) and to the
heuristic used in computing log entries dependencies, which is affected
by the coarse-grained timestamp granularity of the logs included in
our benchmark.
In contrast, MINT, when used as a stand-alone tool on the 
same large logs, does not scale and fares poorly in terms of recall, 
generating very incomplete models.

From a practical perspective, the results achieved by \app lead to a
considerable reduction of false negatives, with a marginal increment
of false positives. For example, for the D15K dataset, MINT generates
(in about two hours) a gFSM that accepts only 52\% of the true
positives (positive logs). In this case, engineers need to
substantially modify the inferred gFSM to accept the remaining 48\% of
positive logs. Instead, for the same dataset, \app generates in about 33
seconds a gFSM that accepts 82\% of the positive logs (and rejects 97\%
of the negative logs). The marginal decrement of the negative logs
correctly dismissed by the gFSM inferred by \app is largely
compensated by (1) a significant reduction of the number of wrongly
rejected positive logs (+30$pp$ in recall), and (2) a substantial
reduction of the execution time (\app is about 222 times faster than
MINT).

In terms of threats to validity, the size of the log files is a
confounding factor that could affect our results (i.e., accuracy and
execution time). We mitigated such a threat by considering seven
datasets with different sizes (ranging from 5K to 35K log entries) and
different sets of system executions.

\section{Related Work}
\label{sec:related-work}

Starting from the seminal work of Biermann and Feldman~\cite{biermann1972synthesis} on the \emph{k-Tail}
algorithm, which is based on the concept of state merging, several approaches 
have been proposed to infer a Finite State Machine (FSM) from execution traces or logs.
\emph{Synoptic}~\cite{Beschastnikh:2011:LEI:2025113.2025151} uses temporal invariants, 
mined from execution traces, to steer the FSM inference process to find models 
that satisfy such invariants; the space of the possible models is then explored 
using a combination of model refinement and coarsening.
\emph{InvariMINT}~\cite{6951474} is an approach enabling the
declarative specification of model inference algorithms in terms of
the types of properties that will be enforced in the inferred model;
the empirical results show that the declarative approach outperforms
procedural implementations of \emph{k-Tail} and
\emph{Synoptic}. Nevertheless, this approach requires prior knowledge
of the properties that should hold on the inferred model; such a
pre-condition cannot be satisfied in contexts (like the one in which
this work is set) where system components are black-boxes and the
knowledge about the system is limited.
Other approaches infer other types of behavioral models that are
richer than an FSM.  \emph{GK-tail+}~\cite{mariani2017gk} infers
guarded FSM (gFSM) by extending the \emph{k-Tail} algorithm and
combining it with Daikon~\cite{ERNST200735} to synthesize constraints
on parameter values; such constraints are represented as guards of the
transitions of the inferred model.
\emph{MINT}~\cite{walkinshaw2016inferring} also infers a gFSM by
combining EDSM (Evidence-Driven State Merging)~\cite{1600197} and data
classifier inference~\cite{mitchell1997machine}. EDSM, based on the
Blue-Fringe algorithm~\cite{10.1007/BFb0054059}, is a popular and
accurate model inference technique, which won the
Abbadingo~\cite{10.1007/BFb0054059} and the StaMinA
competition~\cite{Walkinshaw2013}. Data-classifier inference
identifies patterns or rules between data values of an event and its
subsequent events.  Using data classifiers, the data rules and their
subsequent events are explicitly tied together. \emph{ReHMM}
(Reinforcement learning-based Hidden Markov Modeling)~\cite{Emam:2018:IEP:3208361.3196883}
infers a gFSM extended with transition probabilities, by using a
hybrid technique that combines stochastic modeling and reinforcement
learning. ReHMM is built on top of MINT; differently from the latter,
it uses a specific data classifier (Hidden Markov model) to deal with
transition probabilities.  All the aforementioned approaches cannot
avoid scalability issues due to the intrinsic computational complexity
of inferring FSM-like models; the minimal consistent FSM inference is
NP complete~\cite{GOLD1967447} and all of the practical approaches are
approximation algorithm with polynomial complexity.

Model inference has also been proposed in the context of distributed
and concurrent
systems. \emph{CSight}~\cite{Beschastnikh:2014:IMC:2568225.2568246}
infers a communicating FSM from logs of vector-timestamped concurrent
executions, by mining temporal properties and refining the inferred
model in a way similar to \emph{Synoptic}.
\emph{MSGMiner}~\cite{Kumar:2011:MMS:1985793} is a framework for
mining graph-based models (called Message Sequence Graphs) of
distributed systems; the nodes of this graph correspond to Message
Sequence Chart, whereas the edges are determined using automata
learning techniques. This work has been further
extended~\cite{Kumar:2012:ICL:2337223} to infer (symbolic) class level
specifications.  However, these approaches require the
availability of channel definitions, i.e., which events are used to
send and receive messages among components.

Liu and Dongen~\cite{7849947} uses a \emph{divide and conquer}
strategy, similar to the one in our \app approach, to infer a
system-level, hierarchical process model (in the form of a Petri net
with nested transitions) from the logs of interleaved components, by
leveraging the calling relation between the methods of different
components. This approach assumes the knowledge of the caller and
callee of each component methods; in our case, we do not have this
information and rely on the \emph{leads-to} relation among log
entries, computed from high-level architectural descriptions and
information about the communication events.

One way to tackle the intrinsic scalability issue of (automata-based)
model inference is to rely on distributed computing models, such as
MapReduce~\cite{Dean:2008:MSD:1327452.1327492}, by transforming the
sequential model inference algorithms into their corresponding
distributed version.  In the case of the \emph{k-Tail} algorithm, the
main idea~\cite{wang2016scalable} is to parallelize the algorithm by
dividing the traces into several groups, and then run an instance of
the sequential algorithm on each of them. A more fine-grained
version~\cite{LUO201713} parallelizes both the trace slicing and the
model synthesis steps.  Being based on MapReduce, both approaches
require to encode the data to be exchanged between mappers and
reducers in the form of key-value pairs. This encoding, especially in
the trace slicing step, is application-specific; hence, it cannot be
used in contexts in which the system is treated as a black-box, with
limited information about the data recorded in the log entries.
Furthermore, though the approach can infer a FSM from large logs of
over 100 million events, the distributed model synthesis can be
significantly slower for $k \ge 2$, since the underlying algorithm is
exponential in $k$.
\section{Conclusion}
\label{sec:conclusion}
In this paper, we addressed the scalability problem of inferring the
model of a component-based system from the individual component-level
logs, assuming only limited (and possibly incomplete) knowledge about
the system. Our approach, called \app, first infers a model of each
system component from the corresponding logs; then, it merges the
individual component models together taking into account the
dependencies among components, as reflected in the logs. Our
evaluation, performed on logs from an industrial system, has shown
that \app can process larger logs, is faster, and yields more accurate
models than a state-of-the-art technique.

As part of future work, we plan to refine the heuristics used for
identifying the dependencies of the log entries between multiple components, 
to take into account logs with
imprecise timestamps and out-of-order messages. We also plan to
evaluate \app on different datasets and to integrate it with other
model inference techniques. Finally, we will assess the effectiveness
of the inferred models in software engineering activities, such as
test case generation.

\section*{Acknowledgment}
This work has received funding from the
European Research Council
under the European Union's Horizon 2020
research and innovation programme (grant agreement
No 694277),
from the Luxembourg National Research Fund (FNR)
under grant agreement No C-PPP17/IS/11602677,
and from a research grant by SES.

\bibliographystyle{IEEEtran}
\bibliography{model_inference.bib}

\end{document}